# Probing Near-Field Thermal Emission of Localized Surface Phonons from Silicon Carbide Nanopillars

Saman Zare[1,2], Ramin Pouria[1,2], Philippe K. Chow[3], Tom Tiwald[4], Carl P. Tripp[2,5], and Sheila Edalatpour[1,2*]

[1]*Department of Mechanical Engineering, University of Maine, Orono, Maine 04469, USA*

[2]*Frontier Institute for Research in Sensor Technologies, University of Maine, Orono, Maine 04469, USA*

[3]*Columbia Nano Initiative, Columbia University, New York, New York 10027, USA*

[4]*Ellipsometry Applications Engineering Group, J.A. Woollam Company, Lincoln, Nebraska 68508, USA*

[5]*Department of Chemistry, University of Maine, Orono, Maine 04469, USA*

[*]*Email: sheila.edalatpour@maine.edu*

**ABSTRACT**

Thermal emission of localized surface phonons (LSPhs) from nanostructures of polaritonic materials is a promising mechanism for tuning the spectrum of near-field thermal radiation. Previous studies have theoretically shown that thermal emission of LSPhs results in narrow-band peaks in the near-field spectra, whose spectral locations can be modulated by changing the dimensions of the nanostructure. However, near-field thermal emission of LSPhs has not been experimentally explored yet. In this study, we measure the spectrum of near-field thermal radiation from arrays of 6H-silicon carbide (6H-SiC) nanopillars using an internal-reflection-element based spectroscopy technique. We present an experimental demonstration of thermal emission of the transverse dipole, quadrupole, and octupole, as well as longitudinal monopole from 6H-SiC nanopillars at a near-field distance from the array. We show that the spectral location of the longitudinal monopole and transverse dipole are significantly affected by the near-field coupling between neighboring nanopillars as well as the intercoupling of the nanopillars and the substrate. We also experimentally demonstrate that the spectrum of near-field thermal radiation from 6H-SiC nanopillar arrays can be tuned by varying the dimensions of the nanopillars, providing an opportunity for designing emitters with tailored near-field thermal radiation.

**KEYWORDS:** Near-field thermal radiation, silicon carbide nanopillars, localized surface phonons, spectral tuning, thermal near-field spectroscopy.

**INTRODUCTION**

Thermal radiation in the near-field regime, that is, at sub-wavelength distances from the emitting medium, can exceed the blackbody limit by several orders of magnitude and be quasi-monochromatic. These unique properties of near-field thermal radiation have attracted significant attention for many promising applications, such as power generation using thermophotovoltaic devices,[1,2] near-field photonic cooling,[3] thermal management of electronic devices,[4,5] thermal rectification,[6,7] and imaging.[8,9] Most of these applications require spectral tuning of the near-field thermal radiation. It has been theoretically proposed that thermal radiation spectra in both far-field and near-field distances from the emitting medium can be tailored via using metamaterials engineered at the sub-wavelength scale.[10-47] Among the proposed metamaterials, are nanostructures made of polaritonic materials such as silicon carbide (SiC).[14,16,20-23,30,35,38,45] Polaritonic materials have negative permittivity in their Reststrahlen band bounded by transverse and longitudinal optical phonon frequencies. The Reststrahlen band of polaritonic materials is located in the infrared part of the electromagnetic spectrum,[48] where thermal emission is commonly spectrally located. It is theoretically shown that metamaterials made of nanoparticles of polaritonic materials can emit localized surface phonons (LSPhs) in their Reststrahlen band, causing narrow-band peaks in the near-field spectra.[14,20,49] It is proposed that the spectral location of the thermally emitted LSPhs can be modulated by varying the size and shape of the nanoparticles.[14,20,49] Tuning the spectrum of far-field thermal radiation by engineering the geometry of the polaritonic metamaterials has been demonstrated experimentally.[47,50-60] However, near-field thermal emission of LSPhs from polaritonic metamaterials has not been experimentally demonstrated yet. Indeed, while there have been several experimental studies on total (spectrally-integrated) near-field radiative heat transfer,[1,4,37,61-84] measurements of the spectrum of near-field thermal radiation have been scarce.[9,85-88] A limited number of studies have experimentally explored the near-field response of polaritonic metamaterials to an external illumination using scattering-scanning near-field optical microscopy (sSNOM).[55,89,90] Using sSNOM, an external infrared electromagnetic field is shined to an AFM tip which is located at a sub-wavelength distance from the metamaterials. The backscattered field from the AFM tip is guided to a Fourier transform infrared spectrometer to determine the near-field response of the metamaterials. The spectral measurements using sSNOM provide insight into the near-field response of the metamaterials to an external illumination, but they are dependent on the direction and polarization of the incident electromagnetic wave and do not represent the direct near-field thermal radiation from the metamaterials.

In this paper, we provide measurements of the spectrum of near-field thermal radiation from periodic arrays of 6H-SiC nanopillars. We experimentally demonstrate that thermal emission of LSPhs results in narrow peaks in the near-field spectrum of 6H-SiC nanopillars. In addition, the effect of the nanopillar height on the spectral location of LSPhs is investigated, and it is experimentally shown that the spectral location of LSPhs in the near-field spectrum can be modulated by varying the height of the nanopillars. The spectral measurements provided in this study, which agree with the theoretically predicted spectra, can pave the way for developing tunable emitters for near-field thermal radiation applications.

**RESULTS AND DISCUSSION**

To demonstrate thermal emission of LSPhs in the near-field regime, two periodic arrays of 6H-SiC nanopillars with a similar pitch, *L*, of 1 μm are fabricated using inductively-coupled plasma reactive ion etching (ICP-RIE) (see Methods for fabrication details). Scanning electron microscopy (SEM) images of the fabricated samples (labeled as Sample A and B) are shown in Figs. 1a and 1b, respectively. The fabricated nanopillars have a conical frustum shape. The bottom and top diameters of the nanopillars in Sample A are ~700 nm and ~500 nm, respectively, while the nanopillars in Sample B have a bottom diameter of ~650 nm and a top diameter of ~500 nm. The interpillar spacings, defined as the bottom-to-bottom gap between two adjacent nanopillars, for Samples A and B are ~300 nm and ~350 nm, respectively. The height of the nanopillars in Sample A is ~1 µm, while the nanopillars in Sample B have a larger height of ~1.6 µm. A conical trench is formed around each nanopillar in both samples. The height and width of the trench is ~50 nm in Sample A (see the inset of Fig. 2b), while the dimensions of the trench in Sample B are about half of those in Sample A (see the inset of Fig. 4b).

The spectra of near-field thermal radiation from the fabricated samples are measured using an experimental technique which utilizes an internal reflection element (IRE) for guiding the evanescent waves to a Fourier-transform infrared (FTIR) spectrometer.[88] In this technique, a trapezoidal ZnSe IRE with low infrared losses is brought to a near-field distance from the heated sample. The low losses of the IRE ensure negligible parasitic thermal emission from the IRE. The distance between the sample and the IRE is established using an array of gold nanopillars deposited on the IRE. The thermally-emitted electromagnetic waves with a parallel component of the wavevector, $k_\rho$, in the range of $\sin(20.4°)n_I k_0 < k_\rho < \sin(69.6°)n_I k_0$ (where $n_I$ and $k_0$ are the refractive index of the IRE and the magnitude of the wavevector in the free space, respectively) are coupled to the IRE and are guided to an FTIR spectrometer where their spectra are recorded (see Methods for schematic and details of the experimental setup). The spectra measured for Sample A and a flat 6H-SiC sample are shown in Fig. 2a at a temperature of 150ºC and a separation gap of ~100 nm between the sample and the IRE. The experimental near-field spectra in Fig. 2a are normalized by the far-field thermal radiation measured for a carbon nanotube (CNT) sample, acting as a blackbody thermal emitter at the same temperature. The CNT sample has an emissivity of ~0.97-0.99 in the wavenumber range of interest, that is, 750-1000 cm$^{-1}$ (see Methods for details of emissivity measurement technique). The spectrum of the flat SiC has a peak at 935 cm$^{-1}$ corresponding to thermal emission of propagating surface phonon polaritons in the Reststrahlen band of 6H-SiC (~796-965 cm$^{-1}$). The near-field spectrum of Sample A is quite different from that of the flat surface, and it has five peaks in the considered spectral region. These peaks are located at 777, 843, 876, 910, and 933 cm$^{-1}$.

For comparison with theory, the spectral energy density emitted by Sample A into the free space is calculated using the SCUFF-LDOS tool of the SCUFF-EM electromagnetic solver[91] and is shown in Fig. 2b. A temperature of 150°C is assumed in the simulations, and the near-field energy density is calculated in the free space at a distance of 100 nm above the nanopillars along their central axis. The ordinary dielectric function of 6H-SiC measured using an ellipsometry technique at a close temperature of 200°C is used in the SCUFF-LDOS simulations (see Methods for details of ellipsometry measurements). The near-field spectral energy density is also calculated for a single, free-standing nanopillar (using the SCUFF-LDOS) and a flat 6H-SiC bulk (using scattering matrix approach[92]) at the same temperature and distance and are shown in Fig. 2b. The theoretical spectrum for Sample A also has five dominant peaks corresponding to those experimentally measured. The peaks in the theoretical spectrum are located at 794, 832, 899, 919, and 946 cm$^{-1}$. The measured locations of the peaks are shifted (by up to 23 cm$^{-1}$) in comparison to the theoretical data. Different factors contribute to the observed shift. This shift is partly due to the assumption of an isotropic dielectric function for 6H-SiC when modeling the energy density using SCUFF-LDOS. As mentioned before, the ordinary dielectric function is used in the SCUFF-LDOS simulations. Another factor contributing to the differences between the measured and predicted spectra is related to the effect of the IRE on the captured near-field signal. As mentioned before, the measured spectra only include thermally-emitted electromagnetic waves with $k_\rho$ in the range of $\sin(20.4°)n_I k_0 < k_\rho < \sin(69.6°)n_I k_0$.[88] However, the simulated energy density includes contributions from all thermally emitted electromagnetic waves with any $k_\rho$. Furthermore, the measured electromagnetic waves go through multiple reflections between the sample and the IRE before being captured by the FTIR spectrometer, while the theoretical energy density is simulated in a free space above the samples.[88] The difference between the range of simulated and captured wavevectors and the multiple reflections of thermal field between the IRE and the sample result in broadening and redshift of the measured near-field peaks in comparison with the predicted spectra in the free space.[88]

The peak predicted at 794 cm$^{-1}$ for Sample A is outside the Reststrahlen band of the 6H-SiC and is due to the large value of the imaginary part of the dielectric function at this wavenumber. The peaks predicted at 919 and 946 cm$^{-1}$ are also observed in the theoretical spectrum of the single free-standing nanopillar. To understand the physical origin of these two peaks, the charge density on the surface of the single free-standing nanopillar when illuminated by a $p$-polarized planar wave is simulated using the SCUFF-SCATTER tool of the SCUFF-EM solver.[93] The $k$-vector of the incoming wave lies in the $x$-$z$ plane with an angle of 25° from the central axis of the nanopillar (see Fig. 3a). The simulated surface charge densities, $\rho$, at 919 cm$^{-1}$ and 946 cm$^{-1}$ are plotted in Figs. 3a and 3b, respectively. It is seen from these figures that the peak at 919 cm$^{-1}$ corresponds to thermal emission of the transverse quadruple mode with $m = 2$ ($m$ being the order of the mode), while the peak at 946 cm$^{-1}$ is associated with the transverse octupole mode with a higher order of $m = 4$. Since the spectral locations of these two transverse modes are the same for Sample A and the single free-standing nanopillar, it is concluded that the transverse quadrupole and octupole modes are affected by neither the near-field interactions between neighboring nanopillars nor the presence of the substrate. The near-field spectrum for a single nanopillar also has a peak at 887 cm$^{-1}$ and a shoulder at 911 cm$^{-1}$ (see the close-up view of the shoulder in the inset of Fig. 3e). The surface charge density for the single nanopillar at these two wavenumbers, presented in Figs. 3c and 3d, respectively, shows that the peak at 887 cm$^{-1}$ is due to thermal emission of the longitudinal monopole with $m = 0$, while the shoulder at 911 cm$^{-1}$ is associated with the transverse dipole mode with $m = 1$. For an on-substrate array (i.e., for Sample A), the locations of the longitudinal monopole and transverse dipole are redshifted to 832 and 899 cm$^{-1}$, respectively. To find the factors contributing to this redshift, the spectral energy density above free-standing and on-substrate arrays of nanopillars is simulated in the spectral range of 800-915 cm$^{-1}$ for various array pitches, $L$, ranging from 3 to 1 μm (interpillar spacing of 2300 to 300 nm), as shown in Fig. 3e. For a free-standing array with $L = 3$ μm, where the interpillar spacing is ~3 times the nanopillars' largest diameter, the near-field interactions between neighboring nanopillars are weak and the spectral locations of the longitudinal monopole and transverse dipole are the same as those for a single nanopillar. As the nanopillars get closer to each other with decreasing $L$ to 1.25 and 1 μm, the interactions between neighboring nanopillars become stronger, resulting in a redshift of the longitudinal monopole and transverse dipole modes. In the presence of a substrate, these two modes are further redshifted (the longitudinal mode by up to ~44 cm$^{-1}$ and the transverse dipole by up to ~12 cm$^{-1}$) for all array pitches (see Fig. 3e). The spectral energy density per unit wavevector emitted by Sample A is plotted versus the wavenumber and the in-plane wavevector in the first Brillouin zone (i.e., the Bloch wavevector) in Fig. 3f. The dispersion of the LSPh modes is also shown in this figure. It is seen from Fig. 3f that the longitudinal monopole of the array is

8the single free-standing nanopillar, it is concluded that the transverse quadrupole and octupole modes are affected by neither the near-field interactions between neighboring nanopillars nor the presence of the substrate. The near-field spectrum for a single nanopillar also has a peak at 887 cm$^{-1}$ and a shoulder at 911 cm$^{-1}$ (see the close-up view of the shoulder in the inset of Fig. 3e). The surface charge density for the single nanopillar at these two wavenumbers, presented in Figs. 3c and 3d, respectively, shows that the peak at 887 cm$^{-1}$ is due to thermal emission of the longitudinal monopole with $m = 0$, while the shoulder at 911 cm$^{-1}$ is associated with the transverse dipole mode with $m = 1$. For an on-substrate array (i.e., for Sample A), the locations of the longitudinal monopole and transverse dipole are redshifted to 832 and 899 cm$^{-1}$, respectively. To find the factors contributing to this redshift, the spectral energy density above free-standing and on-substrate arrays of nanopillars is simulated in the spectral range of 800-915 cm$^{-1}$ for various array pitches, $L$, ranging from 3 to 1 μm (interpillar spacing of 2300 to 300 nm), as shown in Fig. 3e. For a free-standing array with $L = 3$ μm, where the interpillar spacing is ~3 times the nanopillars' largest diameter, the near-field interactions between neighboring nanopillars are weak and the spectral locations of the longitudinal monopole and transverse dipole are the same as those for a single nanopillar. As the nanopillars get closer to each other with decreasing $L$ to 1.25 and 1 μm, the interactions between neighboring nanopillars become stronger, resulting in a redshift of the longitudinal monopole and transverse dipole modes. In the presence of a substrate, these two modes are further redshifted (the longitudinal mode by up to ~44 cm$^{-1}$ and the transverse dipole by up to ~12 cm$^{-1}$) for all array pitches (see Fig. 3e). The spectral energy density per unit wavevector emitted by Sample A is plotted versus the wavenumber and the in-plane wavevector in the first Brillouin zone (i.e., the Bloch wavevector) in Fig. 3f. The dispersion of the LSPh modes is also shown in this figure. It is seen from Fig. 3f that the longitudinal monopole of the array is dispersive, while the transverse modes are almost dispersion-less. The dispersion of the longitudinal monopole of the array is affected by the substrate and the array pitch. Figures 3g and 3h show the spectral energy density per unit wavevector for free-standing (Fig. 3g) and on-substrate (Fig. 3h) arrays with three pitches of $L$ = 1, 2, and 3 µm. It is seen from these figures that the longitudinal monopole becomes more dispersive and significantly redshifted in the presence of the substrate. It is also observed that the dispersion of the longitudinal monopole becomes stronger as the array pitch decreases. Figures 3f and 3h further explain the strong spectral redshift of the longitudinal monopole observed for short pitch arrays of nanopillars in the presence of the substrate as discussed above and shown in Fig. 3e.

To study the effect of distance on the spectral locations of the near-field peaks, the spectrum of near-field energy density is calculated for Sample A at four different observation distances of $d$ = 50, 100, 200, and 400 nm, as shown in Fig. 3i. It is seen from this figure that the spectral locations of the peaks are the same for all four distances. However, the relative intensities of the near-field peaks are affected by the observation distance. While the near-field spectra at small distances of $d$ = 50 and 100 nm are dominated by the transverse octupole of the nanopillar array, the longitudinal monopole and the transverse quadrupole of the array become dominant at larger distances of $d$ = 200 and 400 nm. A small distance of $d$ = 100 nm has been considered in this study as the intensity of the collected signal by the FTIR spectrometer is stronger at smaller $d$s due to the exponential decay of the thermal near field with $d$.

The measured near-field spectrum for Sample B is compared with that for Sample A in Fig. 4a. The simulated spectra for both samples are also shown in Fig. 4b. The theoretical near-field spectrum for Sample B has six peaks which are detected in the measurements. The spectral locations of the measured and predicted peaks are specified in Figs. 4a and 4b. The peaks due to the large imaginary part of the dielectric function, transverse quadrupole, and transverse octupole are almost unchanged as the height of the nanopillars increases from 1 μm in Sample A to 1.6 μm in Sample B. To find the physical origin of the peaks predicted at 810 and 875 cm$^{-1}$ for Sample B, the energy densities emitted from a single free-standing nanopillar, a free-standing array of nanopillars with pitch $L$ ($L$ = 1 and 3 μm), as well as an on-substrate nanopillar array with the same pitch $L$ are simulated and presented in Fig. 4c. It is seen that the spectrum of the single nanopillar has two peaks at 859 and 908 cm$^{-1}$. Based on the charge density distributions in Figs. 4d and 4e, these two peaks, respectively, correspond to the longitudinal monopole and transverse dipole modes of the nanopillar which are redshifted on increasing the height of the nanopillar to 1.6 μm. As for Sample A, the spectral locations of these modes are further redshifted for an on-substrate nanopillar array due to the near-field coupling between neighboring pillars and the presence of the substrate (see Fig. 4c). The theoretical near-field spectrum for Sample B also shows a small peak at 903 cm$^{-1}$, which is not observed in the spectra of the single nanopillar and free-standing array, indicating that this peak is caused by the coupling between the nanopillars and the substrate. This peak is detected in the measured spectrum. The observed differences between the measured spectra for Samples A and B, which are in agreement with the theoretical predictions, demonstrate that the spectrum of near-field thermal radiation from nanopillar arrays of polaritonic materials can be modulated by varying the dimensions of the nanopillars.

The near-field spectrum for Sample A is also compared with the spectrum of far-field thermal radiation from this sample as characterized using the spectral emissivity. The spectral, directional emissivity of the sample is theoretically predicted at a temperature of 150ºC and at an angle of 30° with respect to the surface normal using the SCUFF-TRANSMISSION tool of the SCUFF-EM solver[94] and is shown in Fig. 5a. The spectral emissivity of Sample A is also measured at the same temperature and angle and is plotted in Fig. 5b. It is seen that both the predicted and measured emissivity spectra for Sample A are different from the near-field thermal spectra for this sample. More specifically, the longitudinal monopole of the nanopillar array, which is theoretically predicted at 832 cm$^{-1}$, is non-radiative and only exists in the near-field spectrum. Figures 5a and 5b also show that the spectral locations of the near-field and far-field peaks can be different.

## CONCLUSIONS

The spectra of near-field thermal emission from periodic arrays of 6H-SiC frustum-shaped nanopillars with nanoscale interpillar spacings fabricated on a 6H-SiC substrate were experimentally explored. Two periodic arrays of different heights were considered. Near-field thermal emission of the transverse dipole, quadrupole, and octupole as well as the longitudinal monopole from the nanopillar arrays is experimentally demonstrated. It is shown that the spectral locations of the transverse quadrupole and octupole do not change with increasing height of the nanopillars, while the longitudinal monopole and transverse dipole modes of the nanopillars are redshifted as their height increases. Our measured spectra show that near-field thermal emission from nanostructures can be very different from the one for flat surfaces and that the near-field spectra can be modulated by engineering the materials at the sub-wavelength scale.

## METHODS

### Nanopillar Fabrication

To fabricate the SiC nanopillars, 6H-SiC <0001> wafers (MSE Supplies) with a thickness of 430 μm were cleaned using piranha solution and DI water. A 100-nm chromium layer was deposited to be used as a hard etch mask using RF sputtering at a rate of 2.6A/s. The wafer was diced into 1.2×1.2 cm$^2$ square chips after being coated with a protective layer of Shipley S1818 photoresist followed by photoresist stripping and another piranha cleaning step. Each chip was coated with a layer of maN-2403 negative-tone resist with a thickness of 300 nm. A Nanobeam Ltd. nB4 e-beam writer with an acceleration voltage of 80 kV and a beam current of 31 nA was used to pattern a 1×1 cm$^2$ array of circles with thickness, diameter, and spacing of 100, 500, 500 nm, respectively, into the resist. The pattern was developed in an AZ 300 MIF developer and then rinsed with DI water. Then, the chromium circles were defined ICP-RIE in a Cl2/O2 plasma (Oxford PlasmaPro Cobra 100). Piranha solution was then used to remove the residual resist. The SiC nanopillars were defined by another ICP-RIE step using a SF6/O2 plasma with a pressure of 6 mTorr, a forward RIE power of 150 W, and an ICP power of 1500 W. The etch duration determined the height of the pillars. A commercial chromium etchant (Transene CR1A) was used to remove residual chromium hard mask at room temperature overnight. Finally, any residual surface oxides and/or particles were removed from the nanopillars using 50:1 buffered oxide etch solution.

**Near-Field Thermal Radiation Spectroscopy**

The spectra of near-field thermal radiation are measured using an experimental technique utilizing an IRE for guiding the thermally-emitted evanescent waves to an FTIR spectrometer.[88] A schematic of the setup used for implementing this spectroscopy technique is shown in Fig. 6. A trapezoidal zinc selenide IRE (Harrick, EM2122) with a bevel angle of 45° and a refractive index of $n_I \approx 2.4$ is used in this setup. To maintain an air gap between the sample and the IRE, an array of 15×15 cylindrical gold nanopillars with a height of ~100 nm, a diameter of ~4 μm, and a spacing of ~620 μm are deposited on a 1×1 cm$^2$ area on one side of the ZnSe IRE. To deposit the gold pillars, a mask was designed with the L-edit program, and the resulting GDSII file was transferred to the RTS Mann Pattern Generator. The features (i.e., circles) from the GDSII file were stepped and exposed onto a mask blank (Nanofilm, HRC) consisting of a glass plate coated with an 80-nm chrome layer and a 530-nm positive photoresist layer. After exposure, the mask was developed in a MIF300 developer, and the exposed layer (the areas around the circles) was wet-etched using a chromium mask etchant (Transene, CE-5M). The mask was then cleaned with acetone, methanol, and isopropanol. Before depositing the posts, the ZnSe IRE was cleaned with acetone, methanol, and isopropanol and was dried in an oven at 120°C for 10 min. Then, it was coated with a layer of AZ nLOF 2020 photoresist and was spun at 2500 RPM for 30 s using a Laurell resist spinner. Subsequently, it was heated for 1 min at 110°C and allowed to cool down for about 10 min. The IRE and the mask were aligned using a Suss Microtec MA6 aligner, and were exposed to 365 nm UV light at 70 mJ/cm$^2$. After exposure, the IRE was heated again for 1 min at 110°C for the post exposure image reversal bake, and then allowed to cool down for about 10 min. The unexposed circles were removed by dipping the IRE in a MIF300 developer.

The ZnSe IRE was then placed into an AJA International sputtering chamber. In the sputtering chamber, a layer of chrome was first deposited on the crystal to work as an adhesion for the gold posts. The deposition parameters were 120 W of DC power in 3.0 mTorr Argon resulting in a deposition rate of 0.7 Å/s for a total thickness of 20 nm. Immediately following the chrome deposition, a layer of gold was deposited on the crystal. The parameters for the gold deposition were 360 W of RF power in 3.0 mTorr of Argon which resulted in a deposition rate of 3.2 Å/s for a total height of 100 nm. The IRE was removed from the sputtering chamber after deposition and was dipped into AZ Kwik Strip photoresist remover. The stripper and the IRE were placed inside a 200 W ultrasonic bath at a temperature of 60°C to accelerate the removal of the photoresist. The solvent and the IRE were kept in the bath overnight to ensure complete removal of the photoresists.

Finally, the dimensions of the deposited gold posts were measured with a Tencor Alphastep 500 surface profilometer.

To capture the near-field radiation signal, the samples, heated to 150°C using a ceramic heater (Thorlabs, HT24S), are placed on the IRE (covered with the gold posts), and a 100-nm-thick air gap between the sample and the IRE is established. The evanescent waves with $k_\rho < n_I k_0$ emitted from the heated sample are converted into propagating waves inside the IRE. The propagating waves inside the IRE experience total internal reflections at the flat surfaces of the IRE and travel inside the IRE toward the IRE beveled surfaces (see Fig. 6). The waves with $k_\rho$ in the range of $\sin(20.4°)n_I k_0 < k_\rho < \sin(69.6°)n_I k_0$ exit the beveled sides of the IRE and are guided to an FTIR spectrometer (ABB-Bomem MB1552E) using a multiple-reflection horizontal attenuated total reflection accessory (Harrick, HorizonTM). The FTIR spectrometer is equipped with a broadband mercury-cadmium-telluride detector (InfraRed Associates Inc.).

The measured near-field spectra of the fabricated samples are normalized by the far-field thermal emission from a CNT sample at the same temperature as the nanopillar samples acting as a blackbody emitter. A distance of 1 mm is kept between the CNT sample and the IRE for the far-field blackbody measurements. The CNT sample is fabricated using floating catalyst chemical vapor deposition (CVD) of CNTs following the microwave plasma enhance CVD of a silica buffer layer,[95,96] and consists of a 100-μm-thick layer of CNTs on a silicon wafer. Figure 7 shows the measured normal emissivity for the CNT sample at different wavenumbers (see next section for the details of emissivity measurements). As it is observed in Fig. 7, the emissivity of the CNT sample varies between 0.968 and 0.995 in the wavenumber range of 750-1000 cm$^{-1}$ considered in this study.

**Measuring the Spectral Emissivity**

To find the spectral, directional emissivity of a sample, it was heated up to 150°C using a Watlow Firerod cartridge heater. The temperature of the sample is measured using a K-type thermocouple connected to a digital thermometer (OMEGA, HH-52). The sample is mounted on a rotary stage (Stand, Inc., 8MR174-11-20) for directional measurements. The signal due to thermal radiation by the sample at an angle (with respect to the normal axis of the emitting surface of the sample) was collected by the mercury-cadmium-telluride detector (Thermo Fisher Scientific, MCT-A/CdTe) of an FTIR spectrometer (Thermo Fisher Scientific, Nicolet iS50). The signal from a blackbody (ISDC, IR-563), at the same temperature as the sample, was also collected by the spectrometer. The background thermal radiation was recorded when the heater and blackbody were turned off, and it was subtracted from the signals collected for the sample and the blackbody. Finally, the emissivity of the sample was obtained by taking the ratio of the sample and blackbody signals (after the subtraction).

**Measuring the Dielectric Function of 6H-SiC Using Ellipsometry**

The ordinary and extraordinary complex dielectric functions of a single-side polished 6H-SiC c-plane (<0001>) substrate was measured using ellipsometry at a temperature of 200°C. For this purpose, the 6H-SiC substrate was mounted inside of a Linkam TSEL1000 high temperature stage[97] on a J. A. Woollam IR-VASE Mark II ellipsometer.[98] The measurements were performed in the range of 333-5900 cm$^{-1}$ at a 50° angle of incidence. The collected data were modeled using a uniaxial anisotropic substrate model. The phonon modes for both ordinary ($\varepsilon_o$) and extraordinary ($\varepsilon_e$) dielectric responses were modeled using four asymmetric Lorentz oscillators, which account for coupling between modes.[99] Using a procedure similar to Tiwald et al.[100] and Herzinger et al.,[101] various oscillator parameters were optimized to fit the model to the ellipsometric data. The measured $\varepsilon_o$ and $\varepsilon_e$ are plotted in Figs. 8a and 8b, respectively.

**FUNDING**

This work is financially supported by the National Science Foundation under Grant No. CBET-1804360.

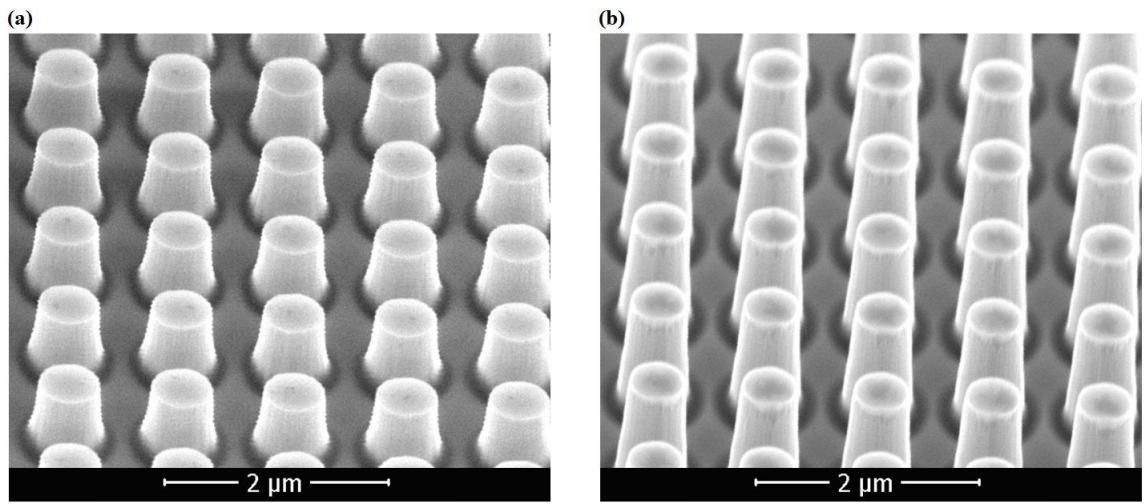

**Figure 1.** SEM images of the fabricated 6H-SiC nanopillar arrays: (a) Sample A and (b) Sample B.

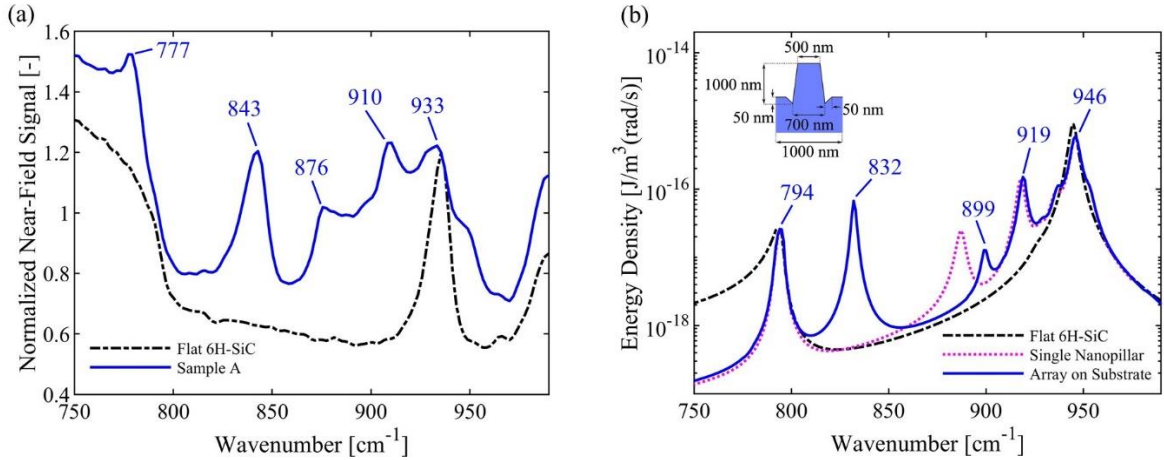

**Figure 2**. (a) Measured near-field thermal radiation spectra for Sample A and a flat 6H-SiC sample at 150°C and a distance of 100 nm above the samples. The near-field spectra are normalized by the far-field thermal emission measured for a CNT sample at the same temperature. (b) Theoretically predicted near-field energy density for Sample A, a flat 6H-SiC sample, and a single free-standing nanopillar at a temperature of 150°C and a distance of 100 nm. The dimensions of the modeled unit cell of Sample A are shown in the inset.

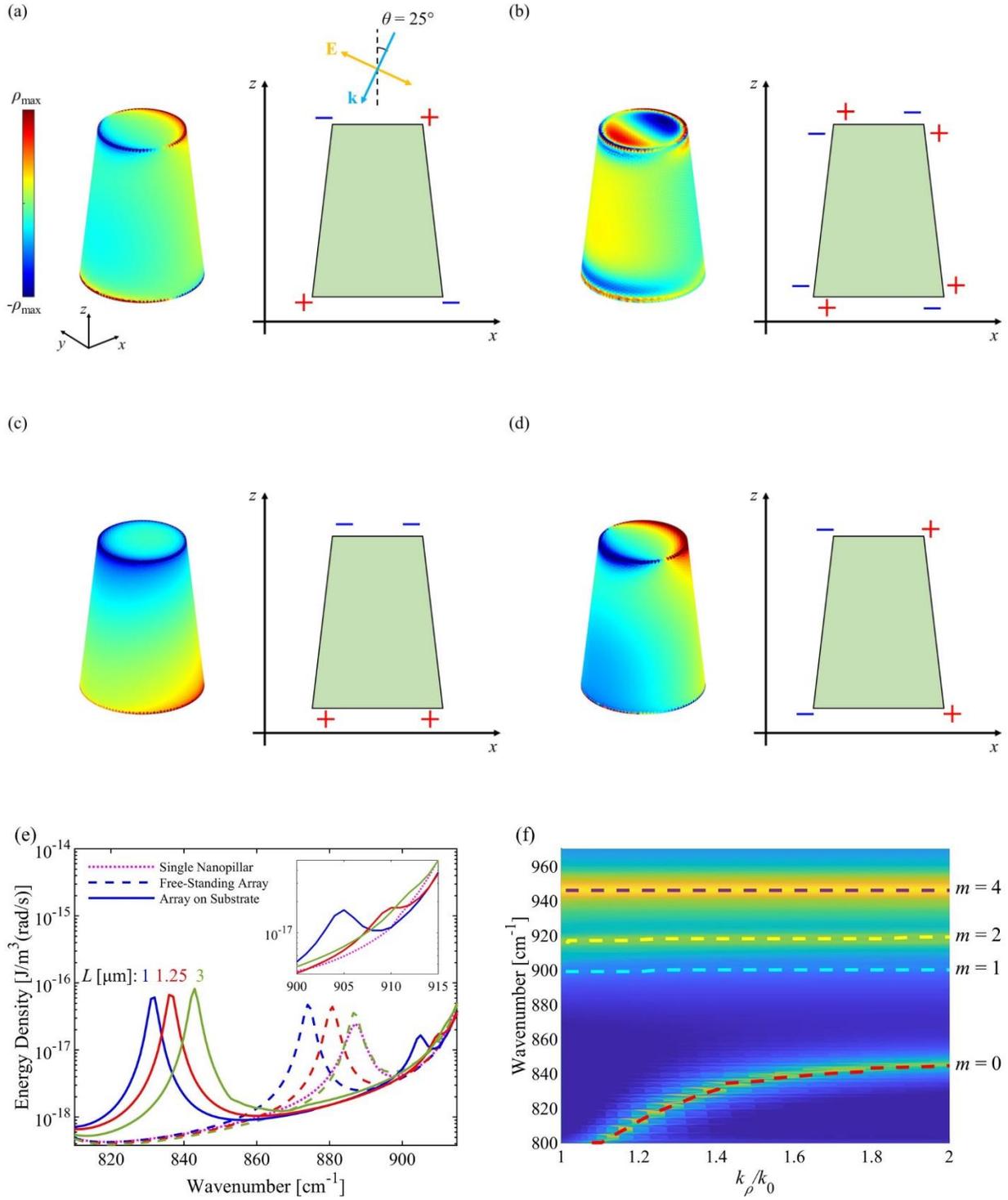

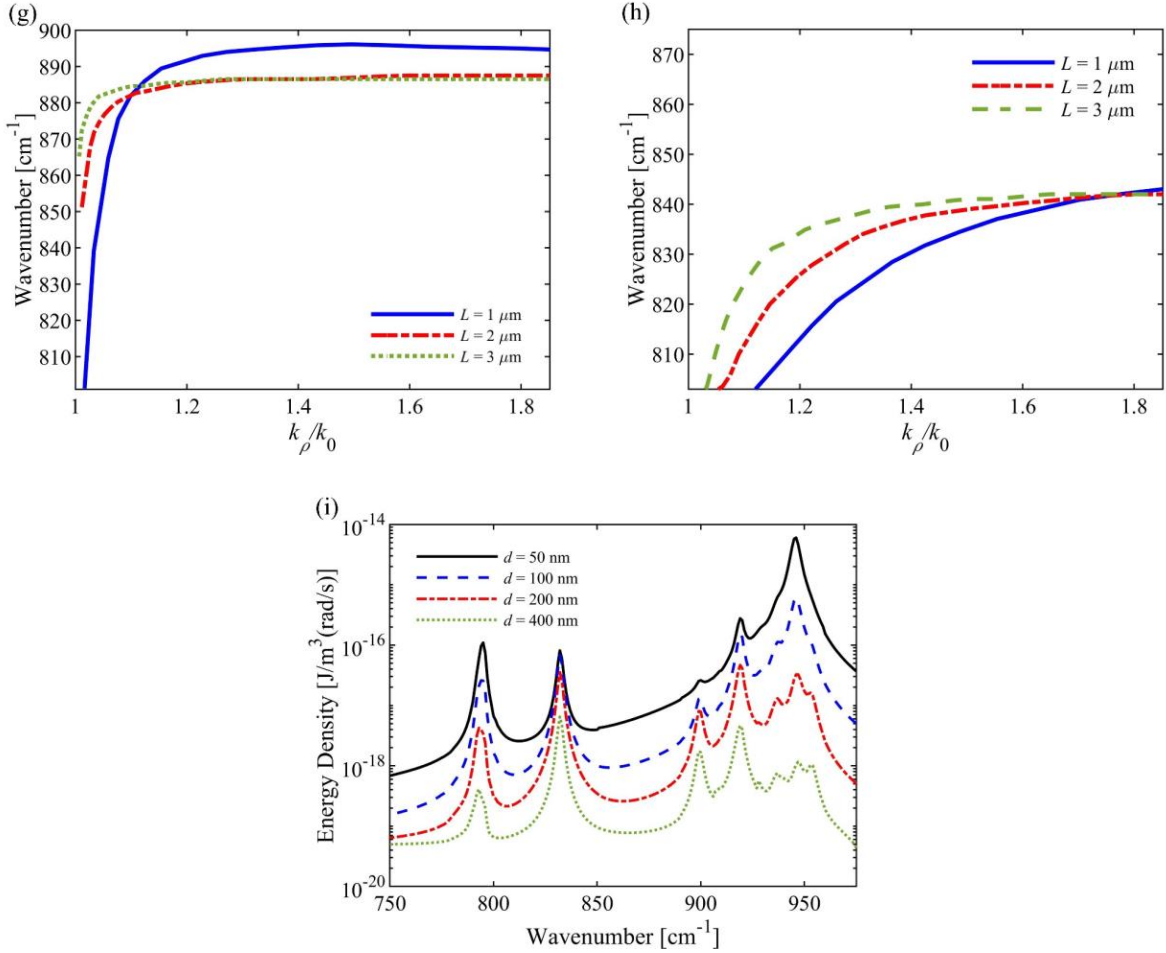

**Figure 3.** Simulated surface charge density, $\rho$, over the surface of a single free-standing nanopillar at (a) 919, (b) 946, (c) 887, and (d) 911 cm$^{-1}$. For each mode, the schematic of the charge distribution in the *x-z* plane is also shown. (e) Predicted near-field energy density emitted by a single nanopillar, a free-standing array of nanopillars, and an on-substrate array of nanopillars for various array pitches, *L*. The inset is a close-up view of the energy density in the spectral range of 900-915 cm$^{-1}$, where transverse dipole mode is emitted. (f) Spectral near-field energy density per unit wavevector emitted by Sample A in the Reststrahlen band of SiC. The dashed lines show the dispersion of the LSPh modes of the sample. (g, h) Dispersion of the longitudinal mode for a (g) free-standing and (h) on-substrate array of nanopillars with various array pitches, *L*. (i) Near-field

energy density predicted for Sample A at a temperature of 150°C and four different observation distances, *d*.

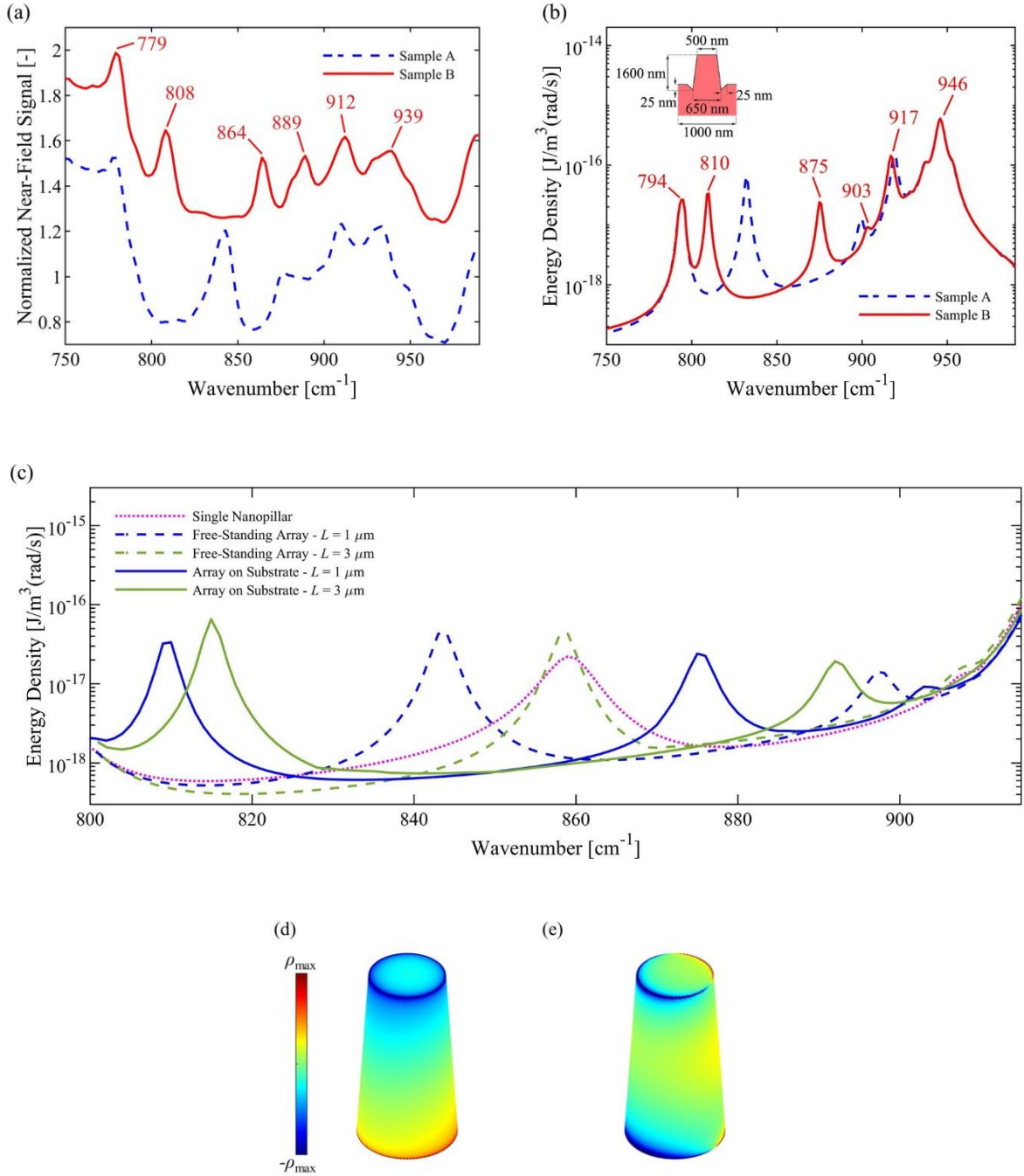

**Figure 4.** (a) Measured spectrum for near-field thermal radiation from Sample B at 150°C and a distance of 100 nm in comparison with that from Sample A. The spectrum of Sample B is offset vertically by 0.35 for clarity. (b) Near-field energy density predicted for Samples A and B at 150°C and a distance of 100 nm. The inset shows the dimensions of the simulated unit cell for Sample B.

(c) Theoretical near-field energy density for a single free standing nanopillar, a free-standing array of nanopillars with pitch $L$ ($L$ = 1 µm and 3 µm), and an on-substrate array of nanopillars with the same pitch. (d,e) Simulated charge density over the surface of a single, free-standing nanopillar corresponding to Sample B at (d) 859 and (e) 908 cm$^{-1}$.

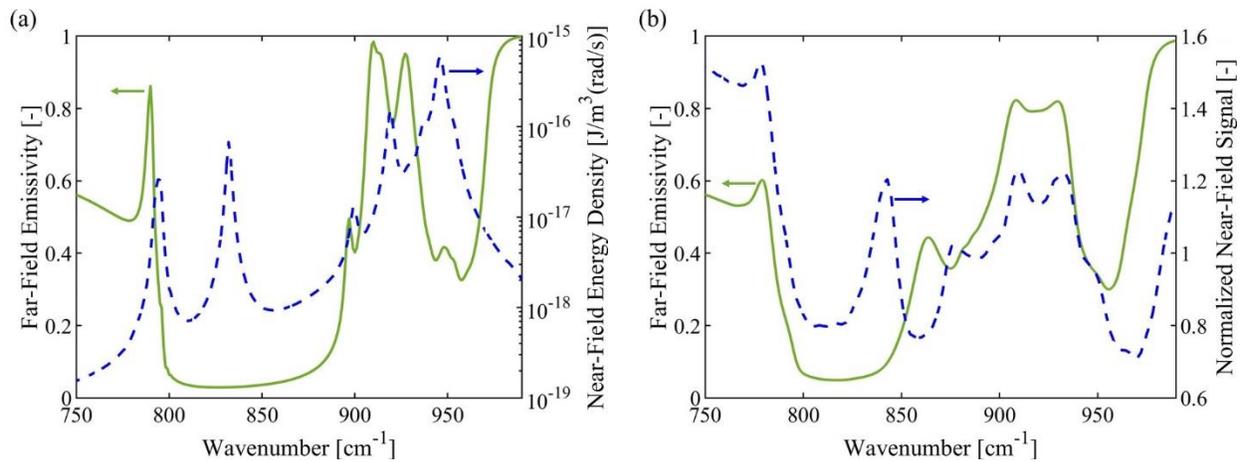

**Figure 5.** (a) Predicted and (b) measured spectral emissivity at an angle of 30° for Sample A in comparison with the spectrum of near-field thermal radiation for this sample.

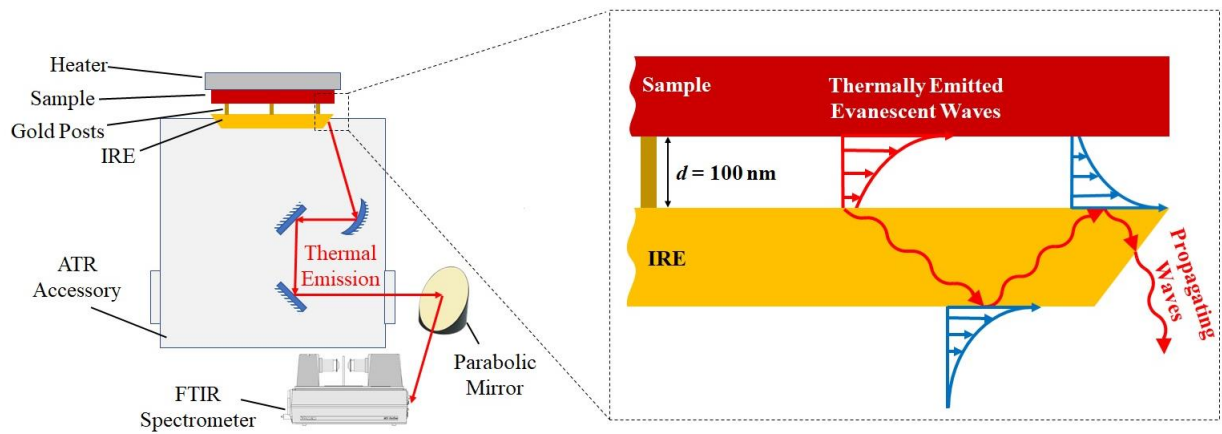

**Figure 6.** Schematic of the experimental setup used for near-field thermal radiation spectroscopy of the fabricated samples. A close-up view of the sample and IRE is also shown.

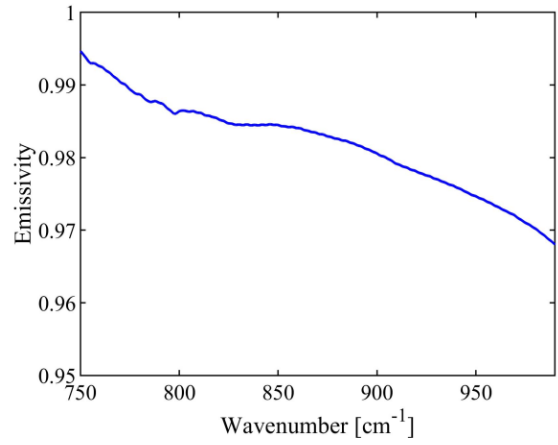

**Figure 7.** Spectral, normal emissivity of the CNT sample measured at 150°C.

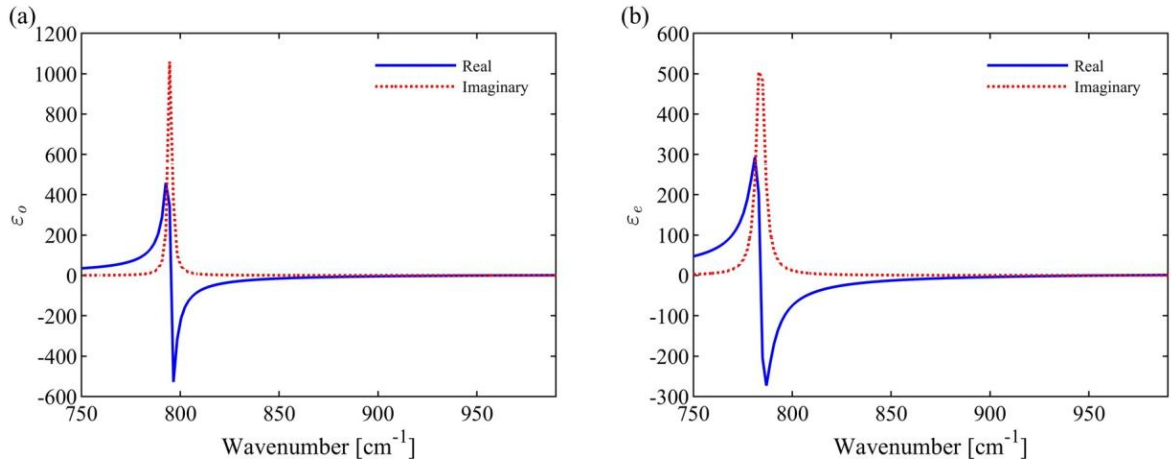

**Figure 8.** (a) Ordinary and (b) extraordinary dielectric functions of a 6H-SiC substrate measured using ellipsometry.